\documentclass[sigconf]{acmart}

\AtBeginDocument{%
  \providecommand\BibTeX{{%
    \normalfont B\kern-0.5em{\scshape i\kern-0.25em b}\kern-0.8em\TeX}}}

\copyrightyear{2021}
\acmYear{2021}
\setcopyright{acmlicensed}\acmConference[WI-IAT '21]{IEEE/WIC/ACM International Conference on Web Intelligence}{December 14--17, 2021}{ESSENDON, VIC, Australia}
\acmBooktitle{IEEE/WIC/ACM International Conference on Web Intelligence (WI-IAT '21), December 14--17, 2021, ESSENDON, VIC, Australia}
\acmPrice{15.00}
\acmDOI{10.1145/3486622.3494026}
\acmISBN{978-1-4503-9115-3/21/12}

\begin{document}

\title{AFFORCE: Actionable Framework for Designing Crowdsourcing Experiences for Older Adults}

\author{Kinga Skorupska}
\email{kinga.skorupska@pja.edu.pl}
\orcid{}
\affiliation{%
  \institution{Polish-Japanese Academy of Information Technology}
  \streetaddress{Koszykowa 86}
  \city{Warsaw}
  \country{Poland}
  \postcode{}
}

\author{Radosław Nielek}
\affiliation{%
  \institution{Polish-Japanese Academy of Information Technology}
  \streetaddress{Koszykowa 86}
  \city{Warsaw}
  \country{Poland}
  \postcode{}
}

\author{Wiesław Kopeć}
\affiliation{%
  \institution{Polish-Japanese Academy of Information Technology}
  \streetaddress{Koszykowa 86}
  \city{Warsaw}
  \country{Poland}
  \postcode{}
}

\renewcommand{\shortauthors}{Skorupska, et al.}

\begin{abstract}
In this article we propose a unique framework for designing attractive and engaging crowdsourcing systems for older adults, which is called AFFORCE (Actionable Framework For Crowdsourcing Experiences). We first categorize and map mitigating factors and barriers to crowdsourcing for older adults to finally discuss, present and combine system elements addressing them into an actionable reference framework. This innovative framework is based on our experience with the design of crowdsourcing systems for older adults in exploratory cases and studies, related work, as well as our and related research at the intersection of older adults' use of ICT, crowdsourcing and citizen science.
\end{abstract}

\begin{CCSXML}
<ccs2012>
   <concept>
       <concept_id>10002951.10003260.10003282.10003296</concept_id>
       <concept_desc>Information systems~Crowdsourcing</concept_desc>
       <concept_significance>500</concept_significance>
       </concept>
   <concept>
       <concept_id>10003456.10010927.10010930.10010932</concept_id>
       <concept_desc>Social and professional topics~Seniors</concept_desc>
       <concept_significance>300</concept_significance>
       </concept>
 </ccs2012>
\end{CCSXML}

\ccsdesc[500]{Information systems~Crowdsourcing}
\ccsdesc[300]{Social and professional topics~Seniors}

\keywords{crowdsourcing, older adults, user experience design, crowdsourcing framework, crowdsourcing system, crowdsourcing experience, active aging}

\begin{teaserfigure}
\includegraphics[width=\textwidth]{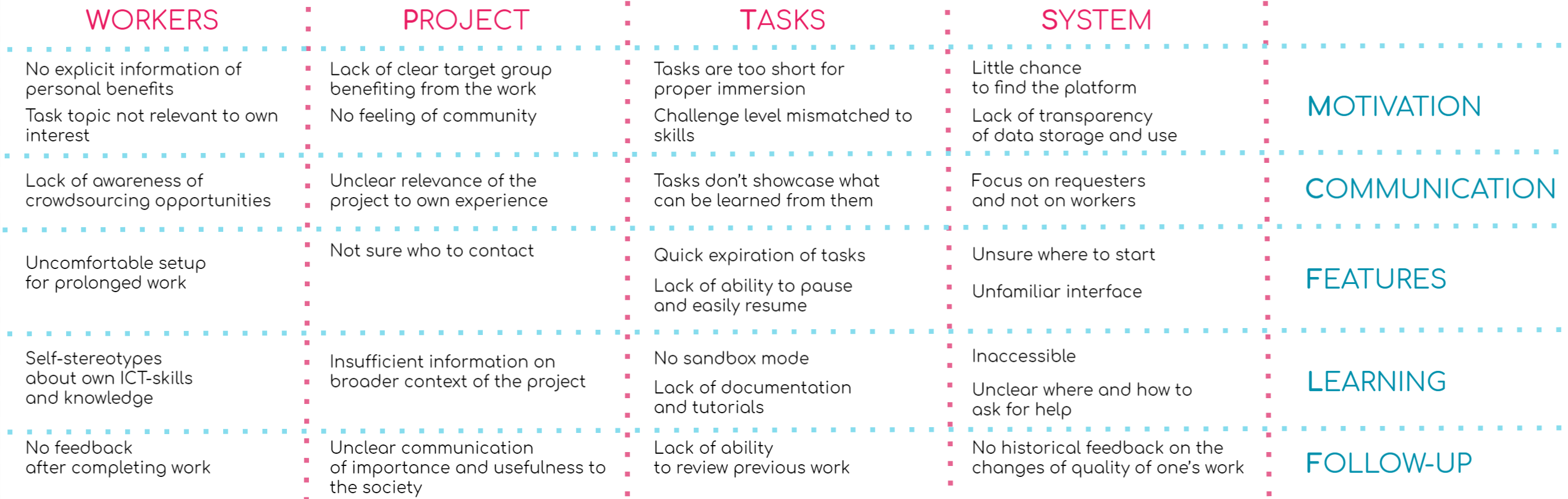}
\caption{A summary of barriers to crowdsourcing which are especially discouraging to older adults. } \label{barriers}
\end{teaserfigure}

\maketitle
\section{Rationale}

%Describe the rationale for the review in the context of what is already known. 

% Although Google Trends places the peak of interest in crowdsourcing sometime in 2014 and the concept is firmly in its second decade there are
Since in 2013 Kittur et al. \cite{kittur2013future} envisioned the future of crowdsourcing, technological advancements made crowdsourcing more efficient, cost-effective and streamlined. Yet, they lag in terms of humanizing the work outside of citizen science projects and efforts driven by volunteers and communities of practice or platforms catering to professional workers. Crowdsourcing platforms offering paid microtask crowd-work place little value in expendable workers, who in turn, may provide poor quality contributions, as the tasks are boring and repetitive \cite{brewerwouldCHI2016}, but the more they do, the better the pay. But even citizen science efforts could up their game to motivate and retain more groups of contributors driven by diverse intrinsic and extrinsic motivations.
Thus, crowdsourcing systems are riddled with shortcomings which inhibit participation for all, but especially for older adults. These barriers, related e.g. to motivation, communication and accessible platform, project and task design are summarized in Fig \ref{barriers} \cite{brewerwouldCHI2016,kobayashimulti2015,crowdolder2020chi,nielek2017,onlinecommunitiesolder2014,volunteerbasedonlinestudies,skorupska2018smarttv,skorupska2020chatbot,sustainablechi2020,kopecspiral2018,mhealtholder2018,accessiblecrowd2015}.

\begin{figure*}
\includegraphics[width=\linewidth]{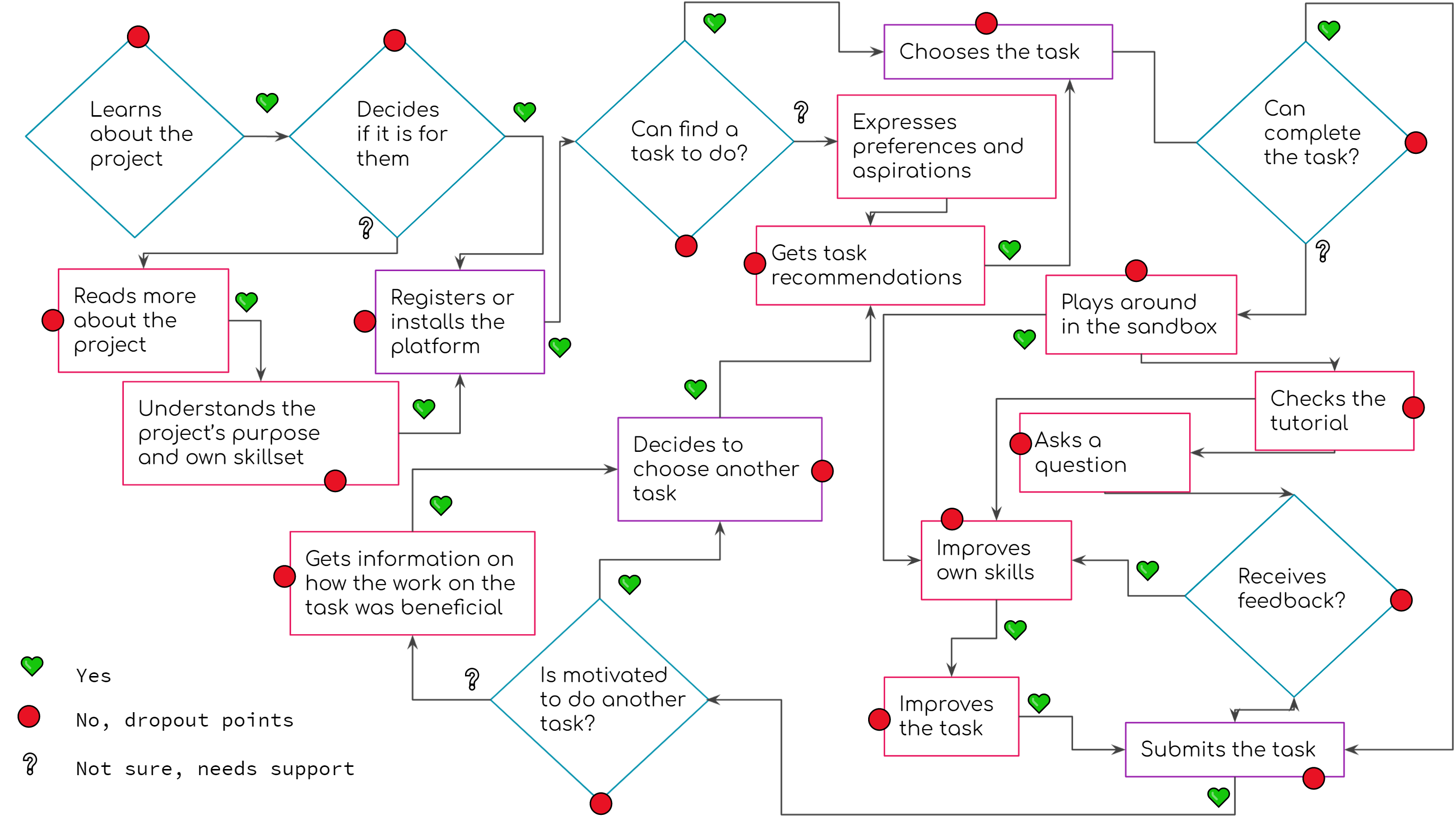}
\caption{A diagram showing the user story on crowdsourcing platforms with indicated possible points of failure based on barriers shown in Fig \ref{barriers}. Potential support elements shown with pink borders.} \label{mapexpe}
\end{figure*}

As the share of older adults, here understood as people aged 65+, especially in western societies, is increasing it more and more important to design remote and scalable ICT-based activities, such as crowdsourcing, that provide them with cognitive challenges, enable positive interaction and active ageing  \cite{skorupska2018smarttv,opensourceolder2014,sustainablechi2020,AmorimMotivation2019BrazilOldCrowd,olderadultssocial2020}. Crowdsourcing projects, on the other hand can benefit from the massive, and largely untapped, potential of older adults' collective intelligence, experience and time \cite{skorupska2018smarttv,knowles2020Conflating,olderadultssocial2020}. Especially that they have proven to be dedicated, diligent and successful contributors \cite{kobayashimulti2015,crowdolder2020chi,skorupska2018smarttv,olderadultssocial2020,zooni2021Interact} who are held back not only by inaccessible task-driven project design, which is incompatible with their prevailing motivations \cite{kittur2013future} and possible age-related cognitive decline in the processing speed and working memory \cite{murman_impact_2015} and the idea of worker expendability \cite{brewerwouldCHI2016,opensourceolder2014,nielek2017,volunteerbasedonlinestudies} but also a lack of support mechanisms \cite{opensourceolder2014,skorupska2018smarttv}, which all result in preventable dropouts, as shown in Fig \ref{mapexpe}.

Therefore, based on our experience with using, designing and developing crowdsourcing platforms, and a keyword-driven literature review done in a form of a mind-map, which allowed us to cluster relevant findings, in this article we envision the way forward to create more engaging, accessible and inclusive experience of crowdwork, encouraging contributions from underrepresented populations \cite{toeppe_conducting_2021,skorupska2020chatbot}. Designing crowdsourcing experiences with older adults in mind, due to the curb-cut effect\footnote{The existence of the curb-cut effect has long been documented in the digital space \cite{hessecurbcut1995} and it differs from universal design, as it first focuses on specific populations who could benefit most from a unique approach.}, will not only improve the experience for them, but also for the general population. To further this goal first we analysed the key barriers to crowdsourcing for older adults, as shown in Fig \ref{barriers} and mapped them over the generalized crowd work user story depicted in Fig \ref{mapexpe} to mark dropout points. Finally, we took Kittur et al.'s proposed framework \cite{kittur2013future} as the starting point for the construction of an actionable framework addressing these dropout hot-spots. Challenges are many, as visible in Fig \ref{barriers}, both related to the accessibility and the required ICT-proficiency \cite{skorupska2018smarttv} and to communication of projects, tasks and their importance. \cite{volunteerbasedonlinestudies,brewerwouldCHI2016} Crowdsourcing systems designed to encourage older adults' contributions ought to not only mitigate barriers but also highlight older adults strong suits: their lifelong experience, skills and patience\cite{skorupska2018smarttv}. They should meet their preferences regarding the experience of contributing, including their individual aspirations.\cite{crowdolder2020chi}

The analysis and elements of the proposed framework are based on our experience with the design of crowdsourcing systems encouraging older adults' contributions in exploratory studies \cite{skorupska2018smarttv,skorupska2019smartTV,skorupska2020chatbot,kopec2017living}, related work, as well as research at the intersection of older adults' use of ICT \cite{orzeszek2017design,balcerzak2017F1,kopecspiral2018} and crowdsourcing \cite{nielek2017wikipedia,nielekcrowdclas2016}. Our resulting framework aims to serve as an actionable reference for designing more sustainable and motivating crowdsourcing experiences that better engage older adults to boost the rates of their participation and help them reap the benefits and protective factors that come from online volunteering, lifelong learning, and staying active longer.

\begin{figure*}
\includegraphics[width=\linewidth]{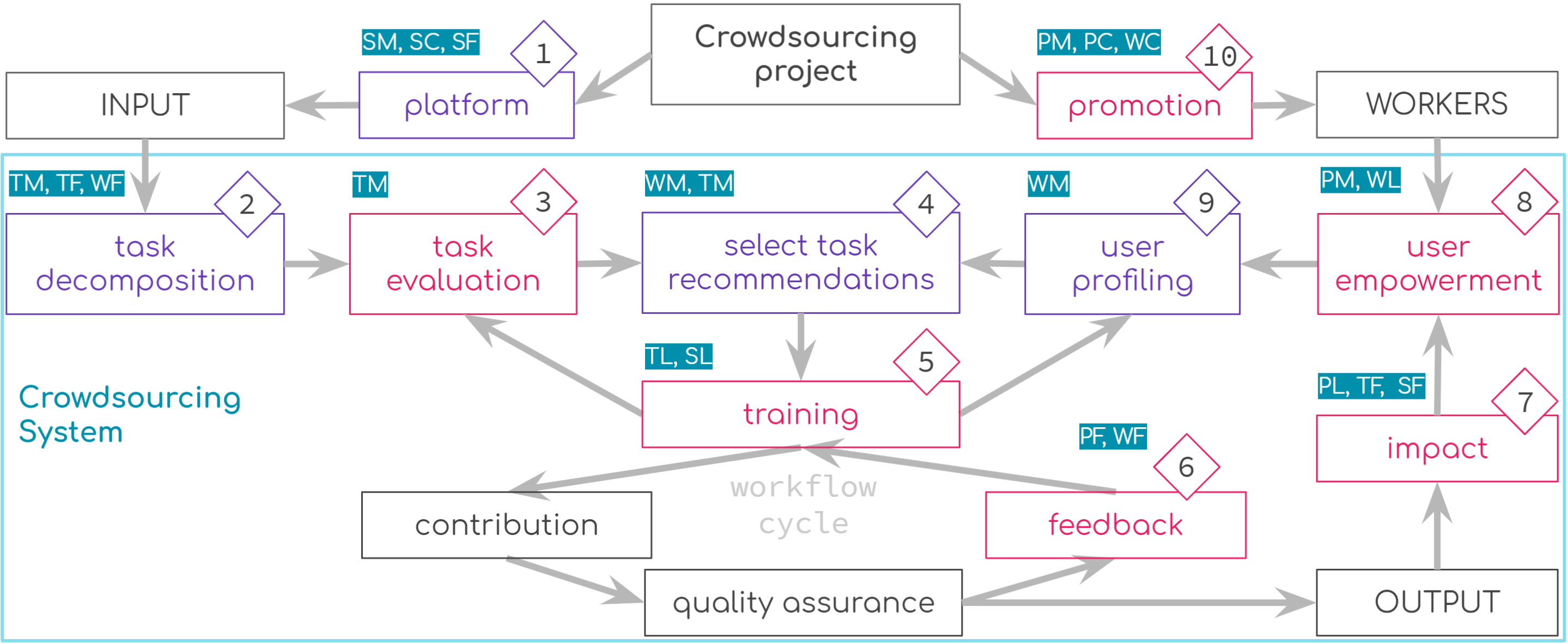}
\caption{Crowdsourcing framework updated with elements important for older adults' engagement - elements in violet are changed, while in pink added in relation to the framework by Kittur et al. \cite{kittur2013future}. Moreover, the barriers addressed by each element are referred to by the first letters of titles of rows and columns in Fig. \ref{barriers}, so e.g. SM stands for System Motivation.} \label{fig1}
\end{figure*}

\section{Proposed Framework}	

What follows is the description of the proposed actionable framework addressing the analysed (Fig. \ref{barriers}) and mapped (Fig. \ref{mapexpe}) barriers. The discussion of framework elements is supported by examples from related research and practice. Special consideration is given to the elements important for designing experiences for older adults.

\subsection{Platform}

When starting a crowdsourcing project the choice of the \textbf{platform (1)} is the project-defining moment, as some of the features described in the framework may not be present in the platform and will need to be supplemented outside of the crowdsourcing system. This is a common problem with the \textbf{feedback (6)} and \textbf{training (5)} module, as well as \textbf{impact (7)} measures and follow-up information on the performance and importance of completed tasks, and quite a challenge, when it comes to \textbf{user empowerment (8)} -- a necessary step when working with older adults using technology \cite{kopecspiral2018,orzeszek2017design}. The \textbf{platform (1)} may also be custom-built, and there was some success with crowdsourcing platforms designed with older adults in mind, especially with co-design \cite{kopecspiral2018,knowles2020Conflating}, which accounted for such aspects as familiarity with the interfaces \cite{Pan2016TheRO}, to help bridge the digital divide, and cater to the characteristic motivation and strengths of older adults \cite{skorupska2018smarttv,crowdolder2020chi}. The importance of this point is clearly visible when evaluating platforms and experiences which were created incrementally and their technical and procedure-driven complexity increased overtime, such as Wikipedia \cite{nielek2017wikipedia}. In the end this complexity constitutes another barrier contributing to the inequality of contributions in terms of less ICT-privileged or experienced user groups which then are underrepresented \cite{skorupska2020chatbot}.

\begin{figure*}
\includegraphics[width=\linewidth]{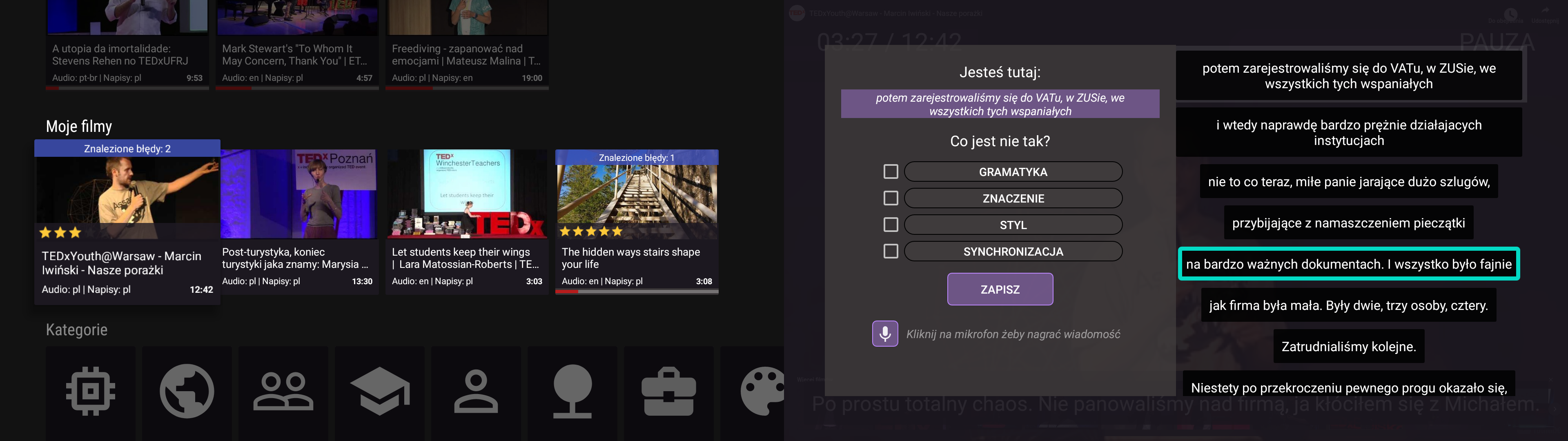}
\caption{Screenshots from a crowdsourcing system designed for older adults and based on the idea of familiar interaction. It also allowed for pausing, viewing the context and free navigating through the task.  Source: https://github.com/manununhez/dreamtv-app.} \label{fig2}
\end{figure*}

Even platform-specific communication makes a big difference. For example, Amazon Mechanical Turk's worker-facing communication focuses on communicating benefits to businesses, calling itself a "crowdsourcing marketplace". On the other hand, the Prolific platform caters to both requesters and participants (contributors) equally and acknowledges the diverse motivations of the crowd with a call to action: "Take part in engaging research, earn cash, and help improve human knowledge" \footnote {Quote and data retrieved on 12.04.2021 from two platforms, commonly used for conducting research: https://www.mturk.com/ and https://www.prolific.co/}. On the other hand, accessibility still plays a role with crowdsourcing platforms active on the market now, small fonts, cryptic pictograms, unfamiliar interfaces and complex procedures make many crowdsourcing platforms inaccessible by default to many groups of users \cite{nielek2017,opensourceolder2014}. This problem, however, is not limited to crowdsourcing platforms or tasks and persists even in applications marketed at older adults \cite{mhealtholder2018}. To alleviate this problem crowdsourcing in a non-computer scenario may bring good results, such as with a remote on a TV screen \cite{skorupska2018smarttv}, making use of familiar modes of interaction, as visible in Fig \ref{fig2} or via a voice interface \cite{skorupska2020chatbot,crowdtasker2020CHI}.

\subsection{Tasks}
The input of the crowdsourcing project should be prepared so that it may be divided into microtasks. However, based on previous research such tasks are often too far removed to understand their purpose \cite{brewerwouldCHI2016} and too short to be meaningful \cite{kittur2013future}. On the other hand, easily relatable tasks, such as tagging historical photos \cite{yu2016productive} or proofreading texts \cite{kobayashiproofreading2013}, are quite encouraging for older adults. Therefore, \textbf{task decomposition (2)} step should strive to produce recognizable work units that the contributors can engage with and become immersed, but can be paused if needed, even for a longer period of time \cite{accessiblecrowd2015} -- perhaps self-specified by the contributor within a pre-set limit, so that users can take a break if they are tired \cite{skorupska2018smarttv} or feel sick \cite{accessiblecrowd2015} and have an overview of the broader context of their work upon return, so that they may again find their footing easily as shown in Fig \ref{fig1} \cite{skorupska2018smarttv}.

These \textbf{mezzo-tasks} should be between microtasks and macrotasks, short enough to complete in a self-defined session length of about 15-30 minutes, but long enough to offer a skill or a piece of knowledge that is valuable to the contributors. Especially older adults are often driven by intrinsic motivation of wanting to learn something interesting \cite{kittur2013future,zooni2021Interact}. For this reason \textbf{task evaluation (3)} is important, so that tasks can be categorized by the skills they utilize, the interest they belong to as well as the level of challenge they pose. Matching the interest, preferred skills and challenge level to the profile of the contributor is necessary for them to stay engaged and to support their development \cite{ordinarychiaging2018}. \textbf{Select task recommendations (4)} ought to be suggested to the contributors to choose from, so that they may exercise their autonomy \cite{skorupska2018smarttv,crowdolder2020chi} and pick work based on their interests, knowledge, skills and energy levels that may change over time \cite{sustainablechi2020}. Very commonly tasks get abandoned \cite{hanabandonment2019} which results in effort and time being wasted, but also in discouragement of the contributors. Crowdsourcing platforms and projects get abandoned along with them. There are many reasons it can happen \cite{opensourceolder2014} including not receiving help, not finding an adequate project at all, or fast enough -- but it can happen at almost any stage of the user journey, as shown in Fig \ref{mapexpe}. This is why \textbf{user profiling (9)} and a shortlist with \textbf{select task recommendations (4)} are so important. Once a user rejects a recommendation an adaptive system could suggest another one, with an increasingly better fit to ensure that they are able to find a task, and successfully complete it -- increasing their well-being \cite{skorupska2020chatbot} and lowering dropout \cite{opensourceolder2014}.

\begin{figure*}
\includegraphics[width=\linewidth]{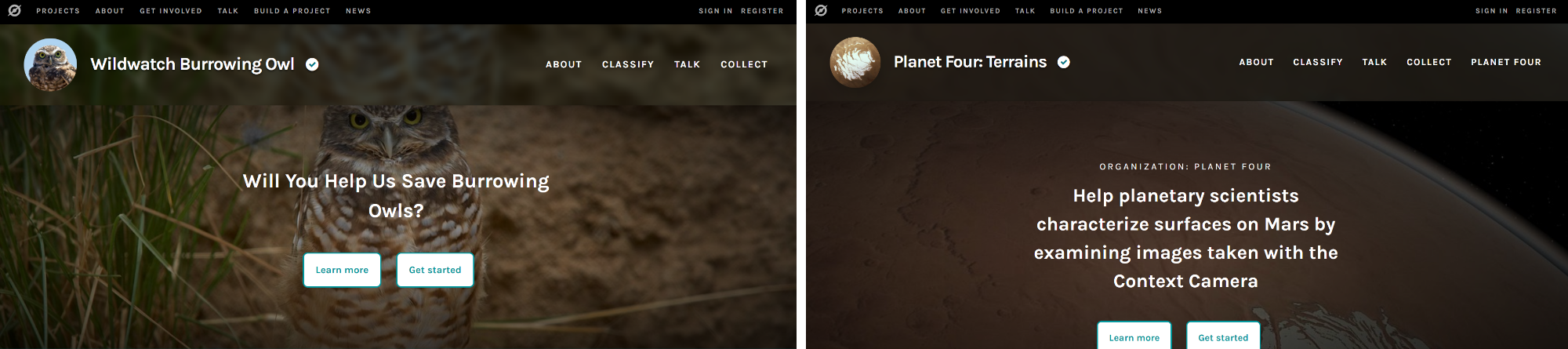}
\caption{Two example calls to action from Zooniverse.org, a citizen science platform -- both appealing to extrinsic motivation. Rephrasing them based on user-profiling, while still keeping true to the nature of the project, could bring better results.} \label{figzooni}
\end{figure*}

\subsection{Empowerment}
Older adults are often apprehensive towards technology and online services, especially if they had little prior contact with them \cite{kopecspiral2018}. In this context \textbf{user profiling (9)} ought to be preceded by \textbf{user empowerment (8)}, so that any self-stereotypes are caught before they affect the profiling. Held stereotypes may affect performance, beliefs about oneself as well as one's skills \cite{knowles2020Conflating}. To address this danger of self-stereotyping and pave way for increased engagement the users ought to learn about crowdsourcing, its purpose and see representative examples. The empowerment step is recommended in many ICT-contexts where older adults are expected to contribute \cite{kopecspiral2018}. They should also be encouraged to trust their own life-course skills, experience and abilities \cite{knowles2020Conflating} which benefit crowdsourcing projects, such as their high crystallized intelligence -- equal to or higher than this of younger adults. Because crystallized intelligence encompasses aspects such as knowledge, understanding vocabulary and concepts and reasoning older adults may be perfect partners for human-in-the-loop systems, which could do the bulk of the crowdsourcing work, with \textbf{quality assurance} left to older adults in areas of their greatest strengths (e.g. language skills, life-course experience, sense-making, individual professional mastery). So, empowerment ought to:

\begin{enumerate}
\item include ICT-empowerment by explaining the importance of crowdsourcing within the ICT practice,
\item incorporate psycho-education to make older adults aware of their strengths which may benefit such projects,
\item and at later stages involve \textbf{training (5)} to do the tasks better,
\item as well as incorporate \textbf{feedback (6)} into the contribution cycle,
\item allow to choose the type of contribution based on individual preferences, \cite{sustainablechi2020}
\item and finally, inform the users of the \textbf{impact (7)} of their contribution, to showcase its societal importance \cite{brewerwouldCHI2016}.
\end{enumerate}

\subsection{Training}
The aspect of \textbf{training (5)} and receiving \textbf{feedback (6)} is important to increase contributors' confidence and skills. Older adults may feel apprehensive of unfamiliar interfaces and new technologies so a sandbox mode, where they can practice without the fear of breaking anything is a requested feature \cite{skorupska2018smarttv}. Another need is related to staying mentally active through challenging oneself \cite{skorupska2018smarttv}, while doing something of societal importance \cite{brewerwouldCHI2016} and learning new things can be such challenge. So, \textbf{task evaluation (3)} should incorporate such tags as "knowledge gained" and "skills practiced" on ladder of different challenge levels, so that users may gain skills to become even better contributors and gain new roles \cite{sustainablechi2020}. The platform ought to also support efforts such as community tutorial generation, either by providing a place to gather collective task insights \cite{microtaskingskillframework} or fostering in-platform communication between users regarding specific tasks, or specific task parts \cite{opensourceolder2014}, but in a positive learning environment, as not to discourage contributors \cite{Halfakerwikidis2013}. Feedback and task performance should stay visible on the contributors' profile for their later reference \cite{volunteerbasedonlinestudies}. The ability to grow one's skill through participation and self-correction is also important, as advancement path and the ability to choose one's role within the project may facilitate sustainable engagement \cite{sustainablechi2020}. Moreover, to increase engagement going beyond online space may be of value, especially to older adults who are isolated and may prefer face-to-face communication \cite{onlinecommunitiesolder2014,crowdolder2020chi}. Such offline supplementary activities can involve training, tutorial writing or content-creation marathons.

\subsection{Profiling and Motivation}
Older adults are a very heterogeneous group which spans decades of life \cite{knowles2020Conflating} during which even successful aging may take very diverse forms in different people. As individual differences are prominent \textbf{user profiling (9)} has to catch diverse users' motivations, aspirations as well as abilities, skills and interests, to enable the system to adapt to its users. Profiling enables to \textbf{select task recommendations (4)} that would be appropriate for each individual, to still grant them freedom to choose their tasks, without causing overload. As small monetary rewards do not sufficiently motivate older adults \cite{crowdolder2020chi,brewerwouldCHI2016} focusing communication on other extrinsic and intrinsic motivators is important. Especially intrinsic motivation plays a greater role for older adults \cite{kittur2013future} so it is necessary to clearly establish aspirations, which especially in western societies most affected by the changing demographic situation, may include the need for entertainment, engaging with their interests or learning new things, and sometimes even staying mentally active to prevent cognitive decline \cite{skorupska2018smarttv,opensourceolder2014,crowdolder2020chi}. Extrinsic motivation, such as the value and usefulness of the task also plays a role, especially if a specific beneficiary is clearly communicated \cite{crowdolder2020chi}. So, the \textbf{impact (7)} that the project has in the real world, including the users' contribution should be made evident \cite{skorupska2020chatbot}. Also, as the energy levels, knowledge, skills, abilities and interests may change over time due to i.e. learning, or natural aging processes \cite{sustainablechi2020} the system should reassess the profile based on task performance, choices and engagement patterns \cite{skorupska2020chatbot}, or even adapt to users' moods and emotions \cite{Shen2017EfficientSI}.
% Among other motivations are feeling useful through conducting socially beneficial activities, such as completing citizen science tasks.

\subsection{Discovery and Communication}
Finally, older adults in general are not familiar with crowdsoucing \cite{brewerwouldCHI2016}, even if they have good ICT skills \cite{zooni2021Interact}. Here, \textbf{promotion (10)} of the crowdsourcing project is necessary in places that older adults already visit online, on platforms they are familiar with. A technology skill and familiarity barrier also exist, but other research suggests that older adults often teach each other tech skills \cite{kowalski2019CHI} if they find them useful and engaging, so it may be enough to promote the project among a group of early adapters, who then take it further. Possibly, they could also recruit each other, as they know the places they frequent best \cite{volunteerbasedonlinestudies}. In communication, it is good to underline such aspects as fun, entertainment or edutainment \cite{skorupska2018smarttv} and the fact that engaging in such activities may support "the increase of self-esteem and social engagement" \cite{AmorimMotivation2019BrazilOldCrowd} along with other benefits of active ageing and project-specific skill and interest highlights.

In general, communication efforts throughout the crowdsourcing platform ought to focus on the evident benefits for older adults themselves and then the society, or, better, a well-specified group within this society, such as children \cite{crowdolder2020chi} or planetary researchers as visible in the message in Figure \ref{figzooni}. Project communication should also be easy, open and two-way, as contributors quite often ask questions about the goals of a project, even in places not designed for this purpose, i.e. comment boxes \cite{volunteerbasedonlinestudies}. Transparent communication is very important \cite{brewerwouldCHI2016} and it should clearly uncover all study and work goals \cite{volunteerbasedonlinestudies} and cover as much context of the work as possible, as limiting information on the project can negatively impact motivation \cite{kobayashimulti2015}. In addition to communication related to recruitment and ongoing issues, such as current projects and open tasks, post-contribution communication is crucial, as older adults especially need to have a proof that their time was purposeful, meaningful and well-spent \cite{olderadultsstrengths2019fomo}.

\section{Conclusions} 

\begin{figure*}
\includegraphics[width=\linewidth]{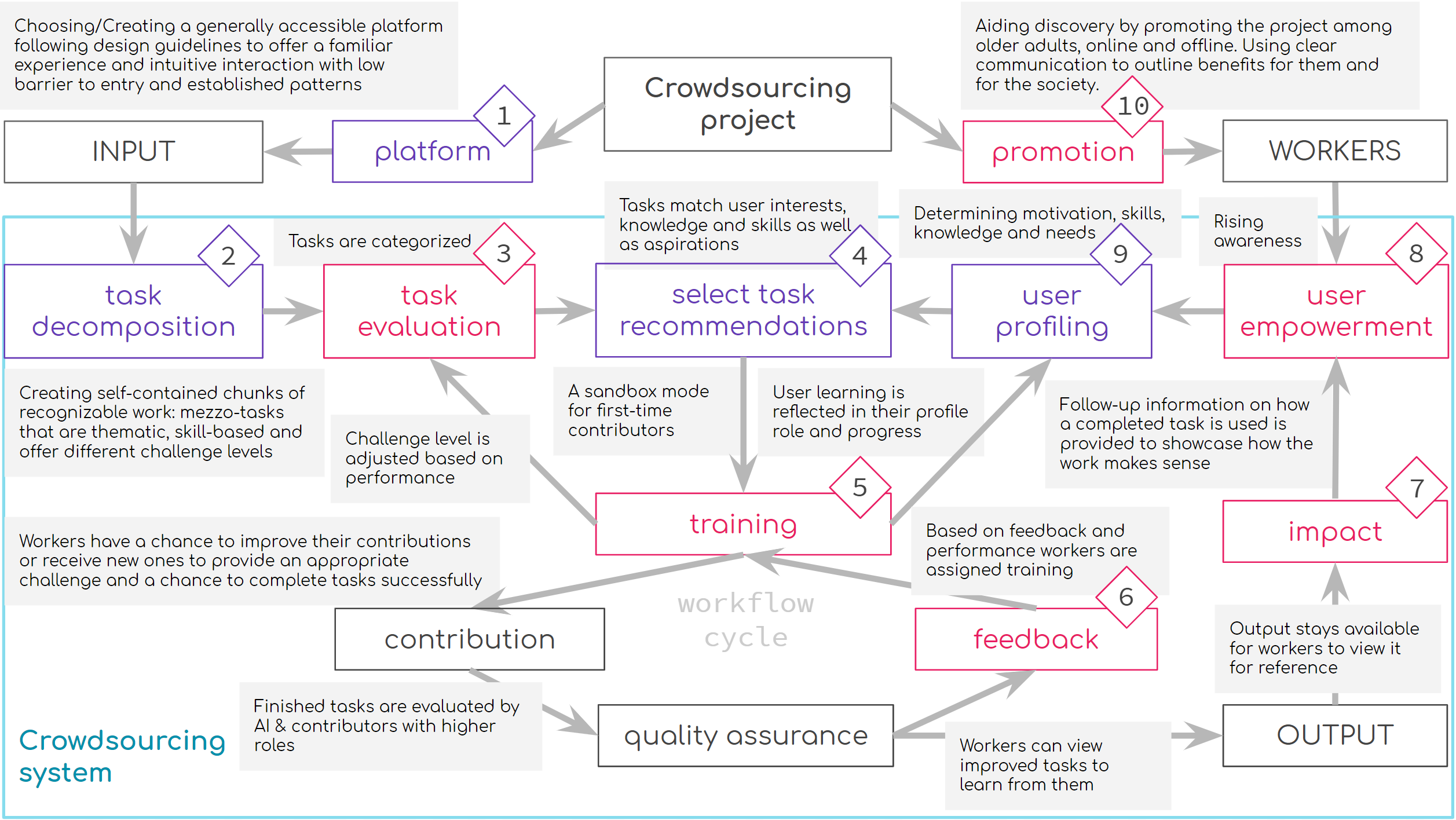}
\caption{A summary of key assumptions of the framework discussed in Section 2.} \label{fullframe}
\end{figure*}

A crowdsourcing project aiming to be successful among older adults ought to be \textbf{well-promoted (10)}, discoverable, available on an accessible \textbf{platform (1)}, consisting of pausable, longer \textbf{self-contained mezzo-tasks (2)}, that are \textbf{evaluated and tagged (3)} based on the interest, skills involved, length and challenge levels. The users should be able to \textbf{select task recommendations (4)} provided based on \textbf{user profiling (9)} done after an \textbf{empowerment (8)} step that levels the playing field for contributors. The users should receive help in the form of tutorials, including \textbf{training (5)} in a sandbox mode. After completing a task, the users should get \textbf{feedback (6)} either from experienced users based on \textbf{quality assurance} or from automated systems, and a chance to \textbf{train (5)} the skills involved as well as information about the \textbf{impact (7)} of their contribution to fulfill both their need to learn, get better and progress in the ranks (intrinsic motivation) and contribute to Social Good (extrinsic motivation) as well as strengthen their motivation to continue contributing. The elements of the framework are summarized in Fig \ref{fullframe} together with key recommended functionalities.

Our hope is that this framework will be used as guidance for other researchers and practitioners delving into the area of creating crowdsourcing systems and experiences for older adults. Moreover, based on the future emerging studies we hope to develop the framework into a more comprehensive guide, towards better inclusion of older adults in ICT-mediated activities.

One limitation of the current crowdsourcing-focused framework is that there are no systems which incorporate all of the elements proposed -- but it is also the motivation behind writing this paper. These proposed elements mitigate the analysed (Fig. \ref{barriers}) and mapped (Fig. \ref{mapexpe}) barriers making crowdsourcing more attractive to some groups of older adults, and due to the curb-cut effect, also to other groups of potential users. Therefore, they should be incorporated into crowdsourcing systems to cater not only to current devoted contributors or task requesters, but also the fringe of the user base -- older adults who soon will be joining the ranks of contributors in greater numbers. 
\\
\textit{This research was partially supported by grants 2018/29/B/HS6/02604 and 2019/35/J/HS6/03166 from the National Science Centre of Poland.}

\bibliographystyle{ACM-Reference-Format}
\balance
\bibliography{bibliography}


\end{document}